\documentclass[a4paper]{article}
\usepackage{amssymb,amsfonts,amsmath}
\usepackage[pdftex]{graphicx}
\pdfoutput=1

\begin{document}

\author{Anna Litvak-Hinenzon and Lewi Stone \\
Biomathematics Unit, Faculty of Life Sciences, Tel Aviv University \\
Emails: annal@post.tau.ac.il, lewi@post.tau.ac.il}
\title{Epidemic Waves, Small Worlds and Targeted Vaccination}
\maketitle

\begin{abstract}
The success of an infectious disease to invade a population is strongly
controlled by the population's specific connectivity structure. Here a
network model is presented as an aid in understanding the role of social
behavior and heterogeneous connectivity in determining the spatio-temporal
patterns of disease dynamics. We explore the controversial origins of
long-term recurrent oscillations believed to be characteristic to diseases
that have a period of temporary immunity after infection. In particular, we
focus on sexually transmitted diseases such as syphilis, where this
controversy is currently under review. Although temporary immunity plays a
key role, it is found that in realistic small-world networks, the social and
sexual behavior of individuals also has great influence in generating
long-term cycles. The model generates circular waves of infection with
unusual spatial dynamics that depend on focal areas that act as pacemakers
in the population. Eradication of the disease can be efficiently achieved by
eliminating the pacemakers with a targeted vaccination scheme. A simple
difference equation model is derived, that captures the infection dynamics
of the network model and gives insights into their origins and their
eradication through vaccination.
\end{abstract}

Developing strategies for controlling the dynamics of epidemics as they
spread through complex population networks is now an issue of great concern 
\cite{AM1991, Riley, EK2002,KE2005, 4gre2002, 4ear2000, Rohani, Eubank,
STD1, STD2}. Future progress depends on gaining a better theoretical
understanding of the spatial dynamics of disease spread, including the
effects of a population's social contact structure and its network topology 
\cite{Riley, EK2002, KE2005, Eubank, Lloyd&May}. Here we show how these
factors control epidemic spread and, in the process, formulate a novel
aggregated targeted vaccination scheme.

We are particularly interested in diseases that confer temporary immunity to
individuals after recovery from infection. This is typical for diseases such
as pertussis, influenza, hRSV and some sexually transmitted diseases (STD's)
as syphilis. In terms of population dynamics, the temporary immunity is
understood to give rise to recurrent epidemic oscillations \cite{gbook} that
can have a period of several years for pertussis \cite{Rohani} and certain
strains of influenza \cite{Flu}, to decadal oscillations in the case of
syphilis \cite{GFG2005, syphLWE98}. In simple terms, the epidemic cycles
arise due to a delayed "SIRS" process in which Susceptible individuals
become Infected, Recover with temporary immunity, but then eventually return
to the Susceptible pool after a time-delay when immunity wears off. The loss
of immunity allows the susceptibles in the population to gradually build up
until sufficient in number to fuel the next disease outbreak.

Grassly et al. \cite{GFG2005} suggested that the oscillations seen in
long-term syphilis data-sets from the US stem from the temporary immunity of
this disease. Their argument is buffered by the fact that gonorrhea, which
lacks temporary immunity, fails to show the same strong cycles in long-term
datasets. This view, however, is controversial and the CDC \cite{CDC2005}
has countered that trends in US syphilis epidemiology follow parallel
changes in population-wide high-risk sexual behavior (see also \cite%
{syphLWE98}). Most likely it is the combined presence of temporary immunity
and social behavior that is responsible for the recurrent waves of syphilis
epidemics. The modeling approach described here allows us to investigate and
assess the impact of these different but important factors.

Complex networks (or graphs) provide an important means for investigating
the effects of social behavior in population models of disease spread.
Individuals are represented as nodes of a graph and edges are placed between
any two individuals should there be an infection route between them \cite%
{EK2002, KE2005, Riley}. A \textit{random} Erdos Renyi network is formed if
there is an equal probability $q$ of a connection between any two
individuals \cite{ErdosRenyi}. A \textit{regular} and tightly clustered
network structure is obtained if an individual is only able to infect
his/her nearest neighbors. The random and clustered-regular graphs might be
considered as two endpoints of a spectrum. Watts and Strogatz \cite{WS1998}
developed a scheme that allows construction of networks that interpolate
anywhere between these two endpoints. This is achieved by introducing a
proportion of $p$ random "short-cuts" between nodes in a regular graph. Only
relatively few short cuts are required ($p<0.1$) to create "small world"
networks that have the often realistic qualities of both a high degree of
clustering, and at the same time relatively high overall network
connectivity introduced by long-range connections (i.e. via short-cuts).

\begin{figure}[tb]
\par
\begin{center}
\includegraphics[width=12cm]{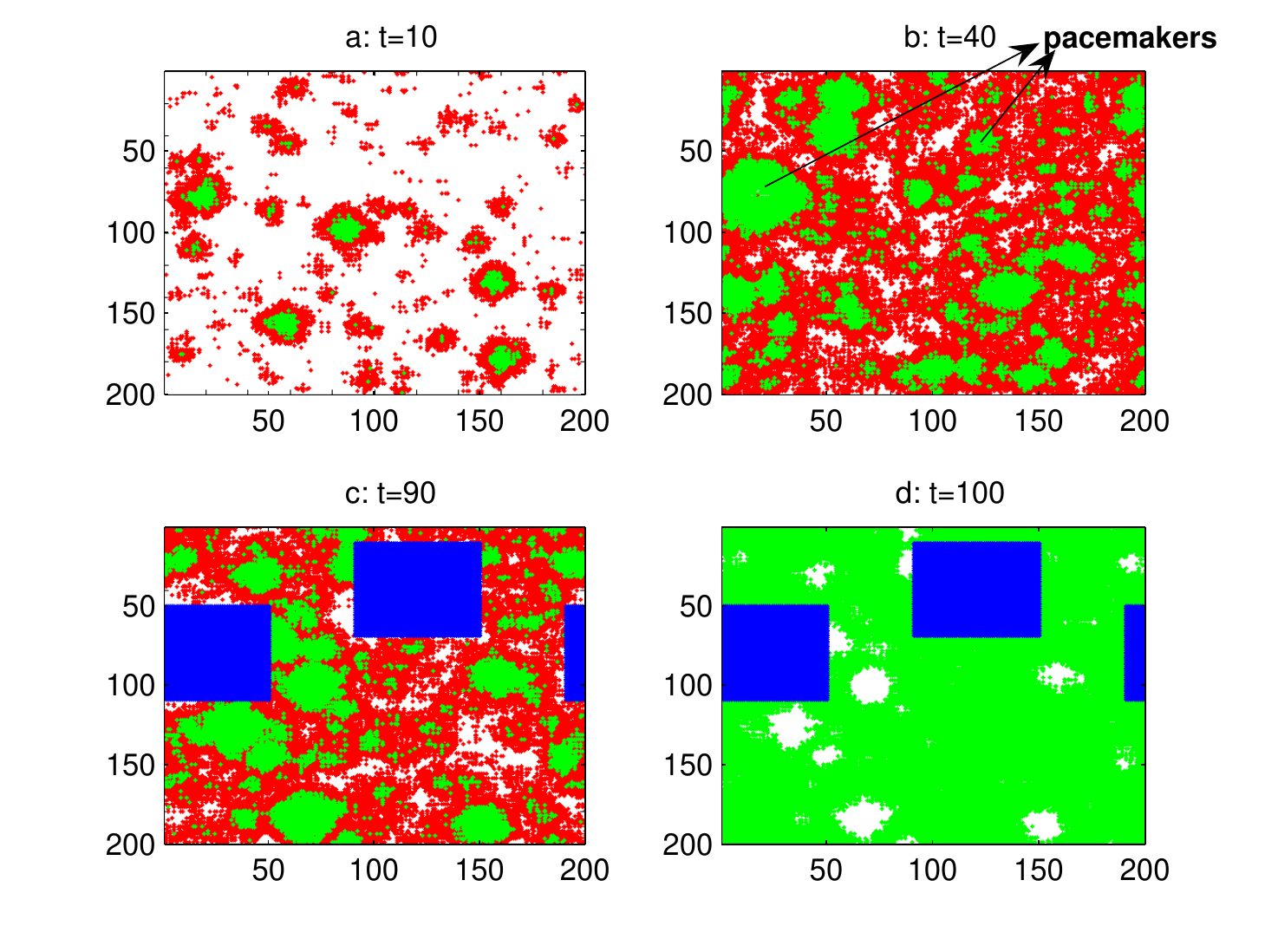} 
\end{center}
\caption{Spatial SIRS model simulation illustrating disease dynamics and
aggregated vaccination. Parameters (as in \protect\cite{KA2001}): $k=3,%
\protect\tau _{I}=4,\protect\tau _{R}=9,p=0.02,q=0.2$. (a) $t=10$. Circular
waves of infected individuals (red) spread through a population of
susceptible (white) and recovered/immune (green) individuals located in a $%
200\times 200$ lattice. (b) $t=40$. Data-analysis has identified two
periodically reappearing pacemaker centers (red rings) of infected
individuals. (c) $t=90$. Aggregated vaccination of all individuals (blue)
located in proximity to the pacemaker centers - comprising only $18\%$ of
the entire population. (d) $t=100$. Disease rapidly brought to extinction
(all red infectives eliminated) in absence of other pacemaker centres.}
\label{f:pacemakers}
\end{figure}

When considering the population dynamics of STD's it is important to take
into account that some individuals spread the disease to a much greater
extent than others. In this way, social behavior and sexual promiscuity
governs the heterogeneity of the contact structure in the population. This
contrasts with standard mean field differential equation models which are
based on the assumption of a "randomly mixing" population and lack a
heterogeneous contact structure. However, there is no unanimous agreement on
how the contact structure of the network should be fixed. Barabasi et al. 
\cite{ScaleFreeScience, ScaleFreeNature} have argued that "scale free"
networks, whose nodes have a power law connectivity distribution, are the
most appropriate for STD's. Lloyd and May \cite{Lloyd&May}, on the other
hand, suggest that such a formulation is unnecessarily exaggerated. We
follow Eames and Keeling \cite{EK2002, KE2005}\ who use a small world model
as a first approximation for STD's. Their model assumes that STD's are
generally transmitted locally but long-range infection pathways exist and
are important in spreading the disease through the population network.
Moreover, the small world approach creates heterogeneity in the connectivity
distribution with most individuals having several connections, but some
being more connected than others.

The small-world formulation is used here to study the spatio-temporal SIRS
dynamics of recurrent diseases with temporal immunity. We first describe the
network model and its spatio-temporal dynamics. For representative
parameters the model exhibits expanding circular waves of infection, some of
which are generated by unusual "pacemaker centers". These we study in
detail; the important role of such "pacemakers" suggests a practical disease
control strategy based on targeted vaccination. We show that by vaccinating
or quarantining the regions around pacemaker centers, the disease can be
eradicated. The vaccination scheme is tested on various more realistic
modifications of the basic model. We then formulate a very simple difference
equation model that captures some of the main features of the heterogeneous
network.

\section{The network SIRS model}

The network model is based on a $2-$dimensional lattice of individuals
(/nodes) whose connectivity $p$ can be preassigned. Each node on the lattice
is occupied and is connected to $k$ nearest neighbors oriented in each of
four directions (North, South, East, West, with diagonal connections
excluded). That is, each node is initially connected to $K=4k$ nearest
neighbors, with $k=3$ unless otherwise specified. The horizontal and the
vertical edges of the lattice are ``glued'' together creating a $2-$torus.
Then, with a probability $p$, each of the $K$ nearest neighbors of each of
the edges in the lattice is randomly rewired to an arbitrary node. These
rewired connections, or ``short-cuts'' \cite{WS1998, KA2001}, may extend to
far regions of the network.

The parameter $p$ controls the population's connectivity structure: $p=0$
corresponds to nearest neighbor contacts only, and where clustering is at
its maximum; small $p$ in the range $0<p<0.1$ corresponds to a ``small
world'' network (each individual has a certain amount of nearest neighbor
contacts + a small proportion of distant contacts, ``short-cuts''); large $%
p>0.4$ is qualitatively equivalent to a randomly mixing population with
minimal clustering \cite{KA2001}. As $p$ is the probability of a short-cut,
it may be viewed as an index of population mobility. Alternatively, it may
be interpreted as an index of social behavior such as sexual promiscuity in
the case of STD's, given the manner in which it controls overall network
connectivity and clustering \cite{WS1998}.

Disease dynamics follow the classical
Susceptible-Infectious-Recovered-Susceptible (SIRS) formulation \cite%
{AM1991, Murray, KA2001, EK2002, GFG2005} with Susceptible individuals ($S$)
having a probability $q$ of becoming infected when linked to an Infected
individual ($I$). Infected individuals eventually recover from the disease
after a fixed time period, $\tau _{I}$, and are conferred temporary
immunity. After a time period of $\tau _R$ time units, immunity wears off
and Recovered individuals ($R$) return once again to the Susceptible pool ($%
S $) closing the SIRS loop.

This is implemented on the network using a $2-$dimensional cellular automata
(CA), SIRS spatial model. At time $t$, an individual at the $(i,j)$'th
location of the lattice has the state $x_{i,j}(t)$ which is either $S,I$ or $%
R$. The model is based on the following transition rules: 
\begin{eqnarray}
&&x_{i,j}(t)\in S\rightarrow \left\{ 
\begin{array}{cc}
x_{i,j}(t+1)\in I & with\quad prob.\quad 1-(1-q)^{k_{\inf }} \\ 
x_{i,j}(t+1)\in S & otherwise%
\end{array}%
\right.  \label{model} \\
&&x_{i,j}(t)\in I\longrightarrow x_{i,j}(t+1)\in I\rightarrow \cdot \cdot
\cdot \longrightarrow x_{i,j}(t+\tau _{I})\in R  \notag \\
&&x_{i,j}(t)\in R\rightarrow x_{i,j}(t+1)\in R\rightarrow \cdot \cdot \cdot
\longrightarrow x_{i,j}(t+\tau_{R})\in S.  \notag
\end{eqnarray}

Infections are transmitted to susceptible individuals with a probability $q$%
, if they are connected to an infective via a nearest neighbor or a
short-cut. Thus the probability that a susceptible becomes infected is $%
1-(1-q)^{k_{\inf }}$, where $k_{\inf }$ is the total number of infected
neighbors of the individual, be they connected via nearest neighbors or via
short-cuts. The proportion of $S(t),I(t)$ and $R(t)$ individuals are
calculated over the lattice and their dynamics are followed as a function of
time.

\begin{figure}[tb]
\par
\begin{center}
\includegraphics[width=12cm]{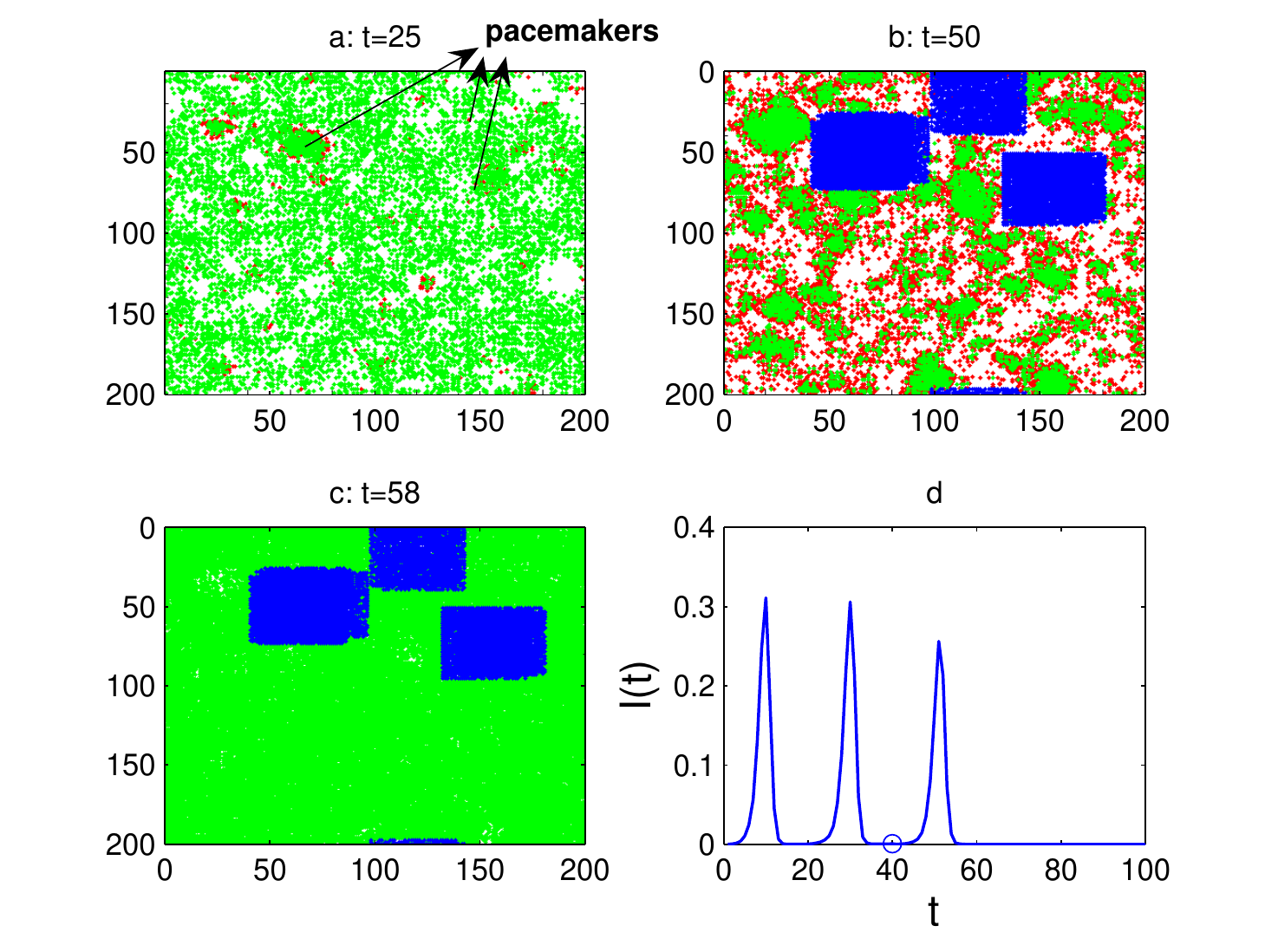} 
\end{center}
\caption{Model simulation of syphilis dynamics (\protect\ref{model}); white
- susceptibles, red - infected, green - recovered, $k=3$, $\protect\tau_{I}=6
$ months, $\protect\tau _{R}=4-8$ years, $p=0.065$, $q=0.36$, $\protect%
\varphi =0.2$ p.y., $\protect\mu =0.02$ p.y. (a) Three pacemakers are seen
at time $t=12.5$ years. (b) Pacemaker areas are vaccinated with random
spread of $86\%$ of each pacemaker area on average, at $t=41$ ($20.5$
years), resulting in vaccinating $14.5\%$ of the population. (c) Disease
eradication at $t=58$ ($28$y.). (d) Time series of the proportion of
infectives per $0.5$y. The period of oscillation is $T\thickapprox 10y$. The
vaccination time is indicated by 'o', after which the disease undergoes
another smaller peak, and reaches complete extinction within less than one
period of the disease ($8.5$ years).}
\label{fig:syphcores}
\end{figure}

To help fix ideas, we focus on two representative parameter settings: (a)
parameters used in a general theoretical model taken from \cite{KA2001}. The
infectious period is fixed at $\tau_{I}=4$ time units and a recovery period
of $\tau_{R}=9$ time units (see simulations in Fig.~\ref{f:pacemakers}); (b)
parameters associated with syphilis epidemics as based closely on the study
of Grassly \textit{et al.} \cite{GFG2005} (see simulations in Fig.~\ref%
{fig:syphcores}). In the latter case $\tau_{I}=1$ time unit, which is taken
to correspond to half a year; the recovery time varies randomly and
uniformly in the range $\tau_{R}\in \{8,9,10,...,16\}$ time units
corresponding to a period of $4-8$ years of immunity. To add realism,
several other features are also incorporated. A birth-death process is
introduced at rate $\mu $, indicating the proportion of births/death per
time step with appropriate network rewiring. In Fig.~\ref{fig:syphcores}, $%
\mu =0.01$ per half-year time step, which is equivalent to a rate of $\mu
=0.02$ per year. Provision is made for the possibility that a proportion $%
\varphi $ of individuals fail to gain immunity after infection (similar to 
\cite{GFG2005}). Thus $\varphi =0$ corresponds to all nodes passing through
an SIRS loop while $\varphi =1$ corresponds to all nodes exhibiting SIS
dynamics (no individual can acquire immunity). In Fig.~\ref{fig:syphcores} a
proportion of $\varphi =0.2$ of the population ($20\%$) gaining no immunity
per year.

\subsection{Recurrent circular waves and pacemaker centers}

We have found that in the small world regime, models of type (\ref{model})
exhibit concentric waves which give rise to unusual ``pacemaker centers.''
Indeed, for $0.001<p<0.15$, the model (\ref{model}) exhibits spatial
oscillations with expanding circular waves of infection traveling through
the lattice (see Fig.~\ref{f:pacemakers}a). Some of these waves are
recurrent, both spatially and temporally. The latter are generated by \emph{%
pacemakers} (Figs.~\ref{f:pacemakers}b and \ref{fig:syphcores}a) that form
at connectivity centers - localized areas denser in short-cuts. The waves
grow in size about the pacemakers as the infection spreads radially. When
infected individuals recover, the interior of the growing wave boundary
becomes a fresh pool of susceptible individuals. At the end of the cycle, a
distant infectee ``short-cuts'' through the network to reinfect the wave's
focal pacemaker, enabling it to perpetuate. The cycle allows recurrent
spatial waves to propagate with a fixed period, $T$. For the syphilis
parameters (see Fig.~\ref{fig:syphcores} legend), $T\thickapprox 10$ years,
as observed in US syphilis datasets \cite{GFG2005, CDC}. Similar spatial
wave patterns had been observed in a number of biological contexts including
epidemiology \cite{GBK2001}, ecology \cite{ecology}, neural networks \cite%
{net} and theoretical studies of excitable systems \cite{GH78}.

These pacemakers follow a pattern formation mechanism. We observe that there
is a minimal necessary amount of short-cuts needed for the creation of a
pacemaker center. In addition, within the small-world range, a pacemaker
wave region has more short-cuts than a non-pacemaker wave region.

The initiation of a pacemaker also requires that the infection is able to
spread both in the horizontal and in the vertical directions. That is, the
development of the infection from an initial state should progress in the
two directions spanning a plane (i.e., in an \emph{X}\ or an \emph{L}
shape). This condition follows the rule that heterogeneities are needed for
creation of spirals in excitable media (see \cite{GH78}). For these reasons,
having an aggregation of short-cuts in a small region enhances the
likelihood of creating a pacemaker. On the other hand, having too large an
aggregation of short-cuts in a localized area of the lattice results in the
opposite effect - disease extinction will occur in this localized area due
to a synchronization effect (see \cite{synch} and below).

\subsection{A Targeted Vaccination Scheme}

The important role of ``pacemaker centers'' suggests a practical control
strategy. We have found that by vaccinating or quarantining the regions
surrounding pacemakers, the disease can be usually brought to a complete
extinction within one period of the disease (in some cases, depending on the
refinement of the vaccination algorithm and specific parameter values, two
vaccination pulses are required). Thus, rather than the conventional scheme
of immunizing some $85\%$ of the population to achieve herd immunity \cite%
{AM1991}, it is only necessary to vaccinate groups enclosing the pacemakers.
This requires vaccination of some $10\%-30\% $ of the population (depending
on the specific application and algorithm refinement; $~\thicksim 20\%$ in
most cases). Fig.~\ref{f:pacemakers} and Fig.~\ref{fig:syphcores} show
spatial snapshots of an infected population upon application of the
vaccination scheme. In the first frames (subfigs. 1a,b and 2a) the
characteristic circular waves of infection (red) are seen. Vaccination
around the pacemakers leads to complete eradication of the disease.
Pacemakers have such large impact on the spatial dynamics that they are
relatively easy to detect using a simple threshold algorithm that identifies
recurring aggregations of infected individuals. Once a pacemaker is
identified, a small region enclosing the pacemaker is marked out and
vaccinated by effectively removing these nodes from the simulation (the blue
rectangles in figures \ref{f:pacemakers}c,d and \ref{fig:syphcores}b,c).

For the theoretical values of \cite{KA2001} used in the simulations of Fig.~%
\ref{f:pacemakers}, the ratio $\tau_{I}/\tau_{R}=4/9$ is relatively large,
hence the pacemakers are large in size and few in number (usually $1-3$). In
some cases, the scheme is able to remove all pacemakers after one
application with vaccinating less than $10\%$ of the population. However, as
the ``pacemakers'' compete with one another, there are cases where only the
main pacemaker(s) is removed and secondary pacemaker(s) may appear in the
next period. Eradication then requires a second application of the
vaccination in the area of the remaining pacemaker(s). As shown in figure %
\ref{f:pacemakers}c,d, in such cases it typically requires vaccinating a
total of some $18\%$ of all individuals over both applications to bring the
disease to total extinction.

For the same model with syphilis parameters, the ratio $\tau_{I}/\tau_{R}$
varies in a simulation within the range $\{1/8,...,1/16\}$. As this ratio is
small and variable, the pacemakers generated are more numerous (usually $2-5$%
) and smaller in size. The additional model realism ($20\%$ of the
population not gaining any immunity, birth/death process) makes the
pacemakers ``less circular'' in shape (see Fig.~\ref{fig:syphcores}a). As a
consequence, it is necessary to vaccinate more areas, although each area is
smaller in size. Nevertheless, the vaccination scheme generally eradicates
the disease in a single application, requiring vaccination of $14\%-22\%$ of
individuals (see Fig.~\ref{fig:syphcores}). As in practice it is difficult
to obtain full coverage when vaccinating an entire population, or even a
specific targeted group, to add realism (and lower vaccination rates) we
vaccinated in Fig.~\ref{fig:syphcores} only an average of $86\%$ of each of
the identified areas, enclosing pacemakers. In more detail, the algorithm
vaccinates up to $98\%$ in the core of the pacemaker, where the high
clustering of repeatedly infected individuals reside and as low as $60\%$
(randomly chosen) in the outskirts of the pacemaker.

It is of interest to examine the effects of simple random vaccination of the
population. For the theoretical parameter values (a) based on \cite{KA2001}
and used in Fig.~\ref{f:pacemakers}, the random scheme is only successful in
eradicating the disease after vaccinating at least $80\%$ of the population.
For the syphilis parameter values, however, the random vaccination threshold
is $43\%$ of the population (this is somewhat similar to the results
presented in \cite{ZK02} for an SIR model). Hence, a further gain can be
achieved by combining random vaccination with the targeted scheme above.
This advantageously reduces the vaccination threshold to some $10\%-15\%$ of
the population for the syphilis parameters. Instead of vaccinating the
entire area surrounding the pacemaker it suffices to randomly vaccinate only 
$60\%$ of the area normally targeted.

In reality, this control scheme may be implemented by vaccination,
quarantine or a targeted education plan, depending on the disease and on the
means available for it's control (for information on efforts to eliminate
syphilis in the US see \cite{syphLWE98} and CDC reports \cite{CDC}). The
main advantage of the vaccination methods proposed here is that they avoid
the usual practice of vaccinating a large proportion of the population. In
addition vaccination is confined solely to relatively small and specified
areas (see figures \ref{f:pacemakers}c,d and \ref{fig:syphcores}b,c). In
practice, it is always preferable to vaccinate as small a group as possible,
as vaccination always carries a risk. Hence, it is advantageous to target
only the relevant groups, already at risk. In the case of syphilis, a
vaccine \cite{syphvac} is still under development, but it is feared to be of
relatively high risk - so if at all, vaccinating only carefully targeted
population already at risk (e.g., in proximity to core groups of highly
active individuals \cite{STDcore, EK2002}) will be desirable if and when
such a vaccine is available. Note, however, that while other works refer to
tracing infected individuals or the most connected individuals for applying
a targeted vaccination scheme (e.g., \cite{ZK02, PSV02, EK2002}), in the
scheme proposed here no contact tracing is required - the pacemaker waves
stem from the SIRS dynamics and the small world structure in a natural and
intrinsic way. Hence, these areas are easily identified as small areas where
the infection appears repeatedly
(see Figs.~\ref{f:pacemakers} and \ref{fig:syphcores}).

\subsection{Synchronization and disease extinction}

Worthy of comment is the model's behavior for larger values of $p$,
typically $p>0.1$, outside the ``small world'' regime, and corresponding to
high levels of sexual promiscuity in the case of STD's. Counterintuitively,
the epidemic consistently dies out abruptly due to the appearance of
large-scale synchronized epidemics \cite{KA2001, synch, ecology} -- a well
known cause of disease extinction. The synchronization manifests with the
formation of large spatial aggregations of infected individuals. Upon
recovery, these infectives gain temporary immunity for a lengthy time
period. Thus the areas that once contained aggregations of infectives,
become exhausted of susceptibles and there is no possibility for an epidemic
to sustain -- it soon dies out.

\begin{figure}[tb]
\par
\begin{center}
\includegraphics[width=8cm]{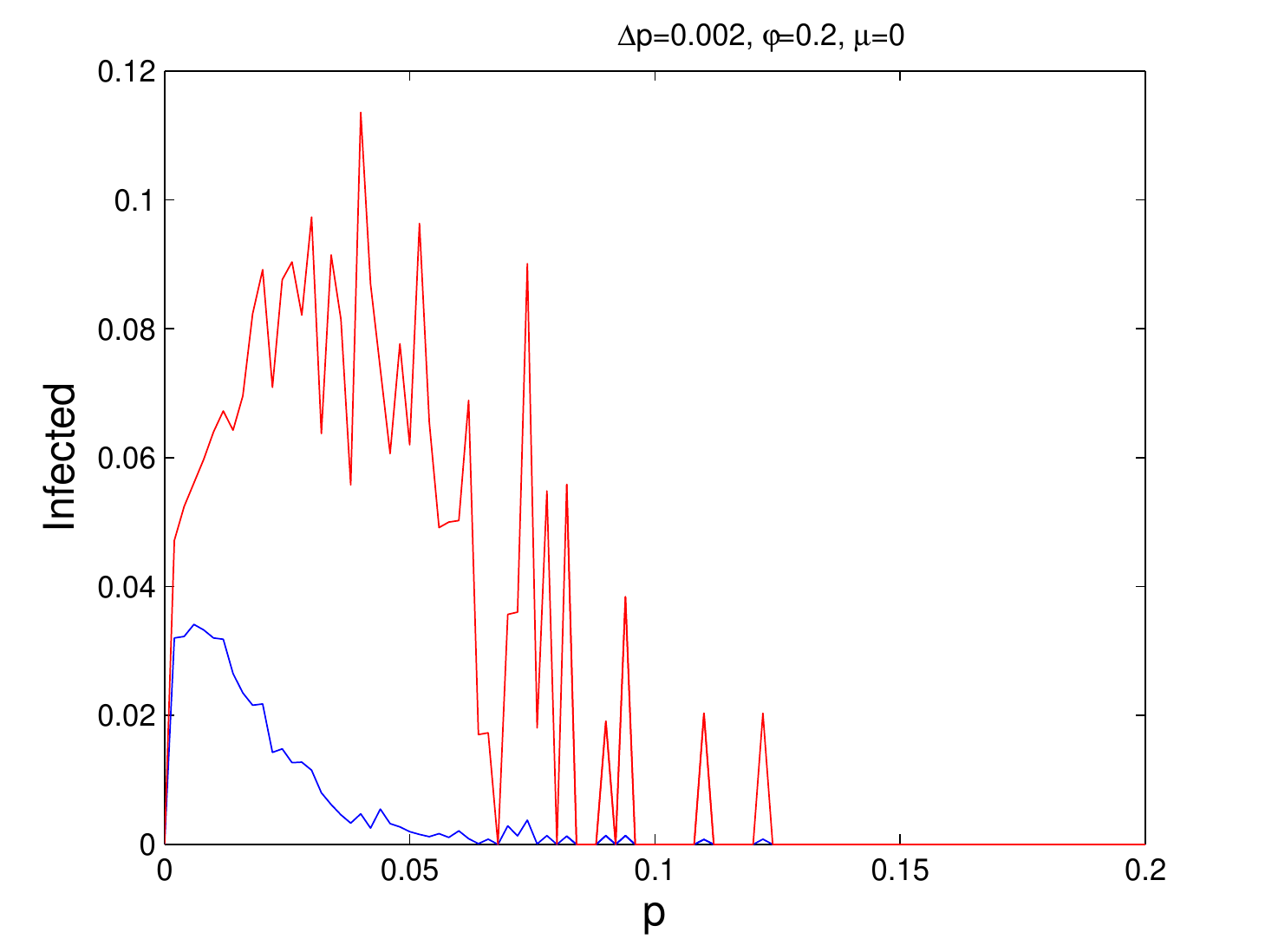} 
\end{center}
\caption{A bifurcation diagram of the proportion of infectives $I$ for
typical syphilis parameter values (as in Fig.2) as a function of shortcuts $%
p $. For small $p$ there is an endemic equilibrium. After a bifurcation
point the dynamics exhibit a limit cycle with radius varying with $p$. The
diagram plots the maximum and minimum values of $I$ on the cycle. After a
second bifurcation value of $p$ the disease goes to extinction (due to
synchronization).}
\label{fig:bif_av}
\end{figure}

Although it might at first seem unusual, the model implies that for society
at large, grand sexual promiscuity has the potential to eliminate STD's such
as syphilis altogether after about two decades of consistent behavior (see
Fig.~\ref{fig:bif_av} and Fig.~\ref{f:de}(c)). 
The same would be true for populations in which
individuals consistently have few proximate sexual partners. The most
conducive conditions for the persistence of such STD's, appears to be the
small world structure similar to the varying manifestation of sexual
promiscuity seen in western society over the last centuries.

\subsection{The impact of short-cuts}

The effect of the parameter $p$ -- the proportion of short-cuts, on the
disease dynamics, may be assessed from the bifurcation diagram in Fig.~\ref%
{fig:bif_av}. The figure plots the range in the number of infectives
(maximum and minimum values) for any given $p$, when the model is run using
the standard syphilis parameters. The following bifurcation scenario takes
place: for $p=0$ the disease goes to extinction; for small $p<0.001$ an
endemic equilibrium is reached in which there is a relatively small
proportion of infectives $0<I^*\ll 1$; for $0.001<p<0.13$ (approximately)
there is a limit cycle of radius depending on $p$ and hence noticeable
oscillations in the number of infectives; for $p\gtrsim 0.13$ the disease
goes to extinction due to a synchronization effect. The ``limit cycle
region'' is the region where pacemaker centers develop and within it lies
the region where the targeted vaccination scheme works very effectively.
This region is an open strip of $p$ values in the ``small world'' regime.
Note that both for $p=0$ and for large $p$ the disease rapidly goes to
extinction. Thus, despite the period of temporary immunity built into this
model, oscillations in $I$ vanish for either very small or relatively large
values of $p$ (i.e., outside the small world regime).

\section{Difference equation model}

We formulate the following difference equation model to help gain insights
into the network model's dynamics. Let $S_{t}$, $I_{t}$ and $R_{t}$ be the
proportion of susceptible, infective and recovered individuals in a large
population at time $t$. Again, let $\tau _{I}$ be the time period an
individual remains infectious and $\tau _{R}$ the period an individual
remains immune. If we assume that $\tau _{I}=1$, as the case for syphilis,
the proportion of recovered individuals can be described by the sum: $\Sigma
_{i=1}^{\tau _{0}-1}I_{t-i}$, and thus: 
\begin{equation}
S_{t}=1-I_{t}-\Sigma _{i=1}^{\tau _{0}-1}I_{t-i},  \label{e:St}
\end{equation}%
where $\tau _{0}=\tau _{I}+\tau _{R}$. The above model formulation is well
known (see e.g., \cite{difeq}), but is extended as follows. We suppose each
individual has on average $K$ connections including those to nearest
neighbors and short-cuts. For the typical individual, denote by $K_{t}^{\inf
K}$ the number of \emph{nearest neighbors} that are infected at time $t$ and
denote by $K_{t}^{\inf p}$ the number of \emph{short-cut links} that point
to infected individuals. Then:

\begin{eqnarray}
K_{t}^{\inf K} &=&I_{t}(1-p)K,  \label{e:KinfK} \\
K_{t}^{\inf p} &=&I_{t}pK.  \label{e:Kinfp}
\end{eqnarray}%
Set $q_{p}$ as the probability of being infected by a short-cut link and set 
$q_{K}$ as the probability of being infected by a nearest neighbor. In
practice $q_{p}>q_{K}$, because of the important role short-cuts play in
spreading the epidemic through the network. For example, simulations of the
lattice model (Eqn.~\ref{model}) under syphilis parameters show that $%
q_{p}\thicksim 4 q_{K}$. For the model parameterized with the theoretical
values taken from \cite{KA2001}, $q_{p}\thicksim 10 q_{K}$. Incorporating
this important observation in the difference equation leads to the following
model of the $SIRS$ dynamics on a complex contact network:

\begin{figure}[tb]
\par
\begin{center}
\includegraphics[width=10cm]{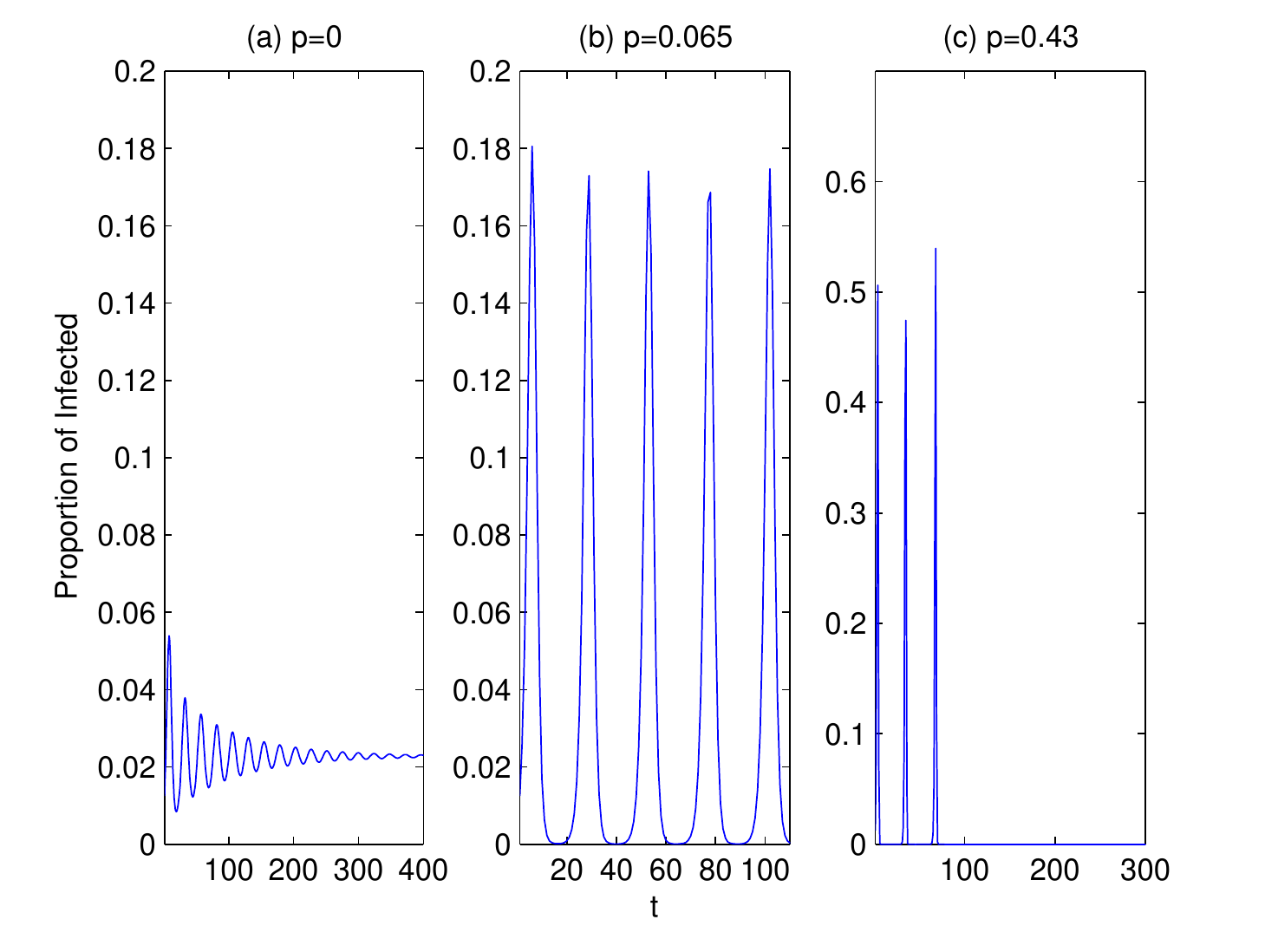} 
\end{center}
\caption{Time series of the difference equation model (\protect\ref{e:Itp1})
for different $p$ values close to bifurcation points; $qK=0.11$, $qp=0.65$, $%
K=12$.}
\label{f:de}
\end{figure}

\begin{equation}
I_{t+1}=(1-(1-q_{K})^{K_{t}^{\inf K}}(1-q_{p})^{K_{t}^{\inf p}})S_{t},
\label{e:Itp1}
\end{equation}%
where $q_{K}<q_{p}\leq 1$. Note that equation (\ref{e:Itp1}) captures both
the time delay dynamics resulting from the temporary immunity of the disease
and the social effects of the short-cuts.


The dynamics of the model (\ref{e:Itp1}) depend on $p$ in a manner that is
very similar to the more complex contact network model (\ref{model}). Fig.~%
\ref{f:de} shows results based on setting $q_{K}=0.11$ and $q_{p}=0.65$. For 
$p=0$, a small endemic equilibrium is reached (see Fig.~\ref{f:de}(a)), for $%
p$ in the small world range sustained oscillations arise (see Fig.4(b))~\ref%
{f:de}(b)), and for large $p$ the disease is eradicated (see Fig.~\ref{f:de}%
(c)) due to a synchronization effect. Comparing figures \ref{fig:syphcores}d
and 4(b) \ref{f:de}(b) -- one can see exactly the same type of dynamics with
similar period of about $10$ years, as observed in syphilis datasets \cite%
{CDC, GFG2005}.

\subsection{Stability analysis}

The equilibrium solutions of model (\ref{e:Itp1}) are found by solving the
following equation for $I^{\ast }$: 
\begin{equation}
I^{\ast }=(1-(1-q_{K})^{I^{\ast }(1-p)K}(1-q_{p})^{I^{\ast }pK})(1-\tau
_{0}I^{\ast }).  \label{e:defp}
\end{equation}%
It is easily seen that the infection-free equilibrium $I^{\ast }=0$ is
always a solution. A stability analysis (based on linearizion of Eqn.~\ref%
{e:Itp1}) reveals that the infection-free equilibrium is stable when the
following inequality holds: 
\begin{equation}
R_{0}=-K[{(1-p)\ln (1-q_{K})+p\ln (1-q_{p})}]\leq 1.  \label{e:bifcrv0}
\end{equation}

Hence, for the given model parameters and relevant $K$ values ($K=8$ or $12$%
), the infection free equilibrium is stable either only for extremely small $%
p$ values ($K=8$), or never ($K=12$). This is visualized in Fig.~\ref%
{f:0st_Hopf}(a) which plots the bifurcation curve of the infection free
equilibrium as a function of the parameters $K$ and $p$. The infection free
equilibrium is stable for all values of parameters below the (lower solid)
curve where $R_{0}<1$ and unstable otherwise since $R_{0}>1$. For $K\leq 4$,
the infection free equilibrium is stable in a wide strip of the parameter
plane, containing the small world regime (as visible in Fig.5a).

\begin{figure}[tb]
\par
\begin{center}
\includegraphics[width=10.2cm]{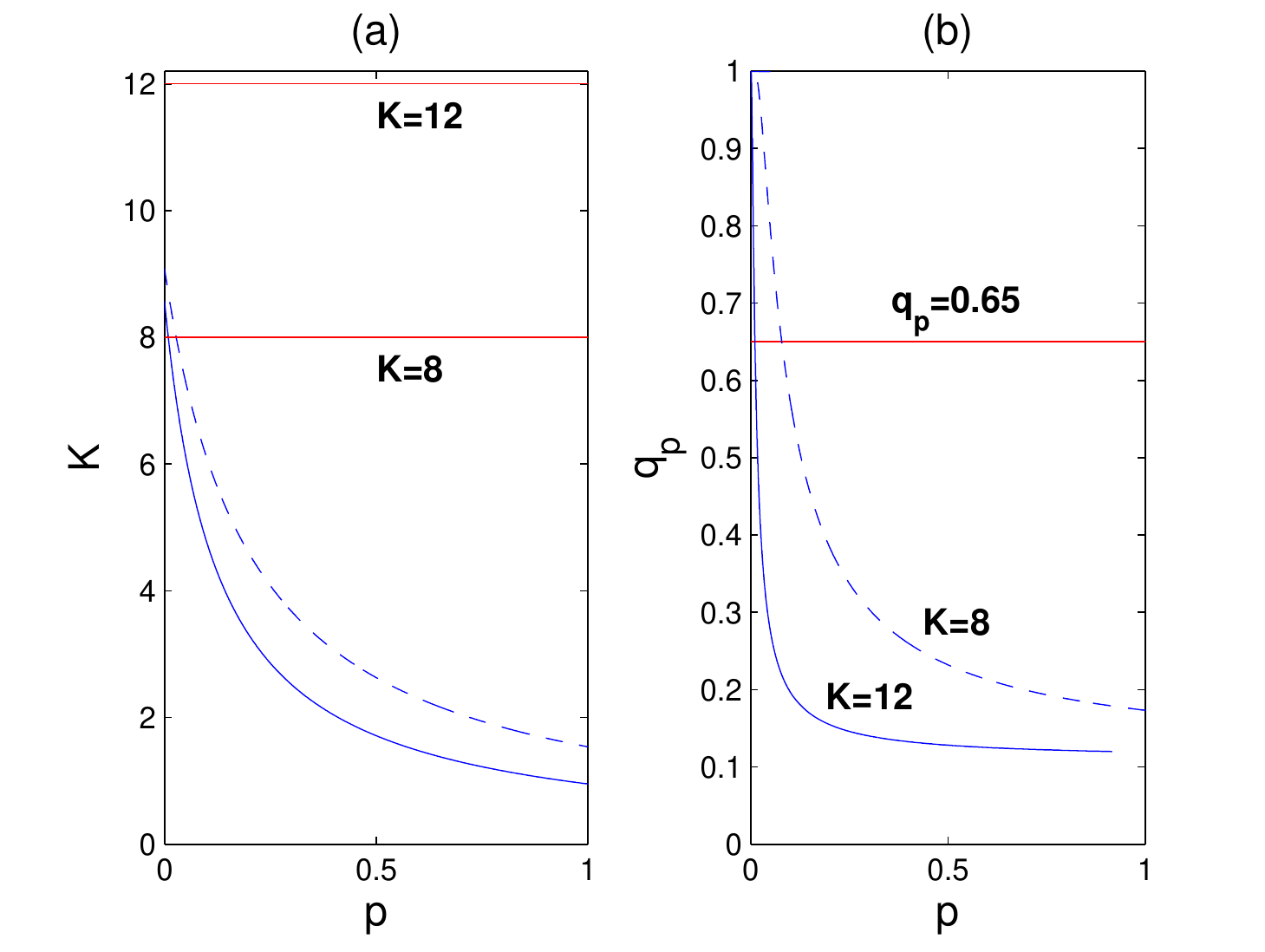} 
\end{center}
\caption{Bifurcation diagrams of the difference equation [\protect\ref%
{e:Itp1}]; $\protect\tau_0=11$, $q_K=0.11$. (a) $q_p=0.65$. Infection free
equilibrium $(I^*=0)$ is stable for parameters below solid curve along which 
$R_0=1$. Dashed curve is the approximation based on estimating $R_0$ (see
text). (b) Hopf bifurcation curves. Endemic equilibrium is stable for
parameters below indicated curves and a stable limit cycle is born when
these curves are crossed (solid curve: $K=12$, dashed curve: $K=8$).}
\label{f:0st_Hopf}
\end{figure}

A similar result is obtained by estimating the reproductive number $R_{0}$,
approximating it as the average number of secondary infectives produced by a
typical infective individual in a sea of susceptibles. In the case here,
where a single node may infect only those nodes it is linked to, the number
of secondary infections may be approximated as: 
\begin{equation}
R_{0}\approx K(q_{K}(1-p)+q_{p}p).  \label{e:R0}
\end{equation}
The above estimate gives a good approximation of the exact condition (\ref%
{e:bifcrv0}) as seen in in Fig.~\ref{f:0st_Hopf}(a) (dashed curve).

For most parameter values, the difference equation (\ref{e:Itp1}) has a
second endemic equilibrium in which $I^{\ast }>0$. The bifurcation curves
describing it's stability in the $(p,q_{p})$ parameter plane are shown in
Fig.~\ref{f:0st_Hopf}(b), for $K=8$ (upper curve) and $K=12$ (lower curve).
This curve indicates a Hopf bifurcation in the $(p,q_{p})$ plane, where all
other parameters are kept fixed. See Appendix1 for technical details on its
calculation. The endemic equilibrium exists for all parameter values for
which the Hopf bifurcation curve exists. Below the bifurcation curve the
endemic equilibrium is stable and is unstable above it, where a stable limit
cycle exists. The stable limit cycle thus coexists with the two unstable
equilibria -- infection free and endemic.

Hence, the bifurcation scenario, for syphilis parameter values, for $K=8$,
is as follows. For $0\leq p\leq 0.0091$, the infection free equilibrium is
stable and the disease goes extinct; for $0.0091<p\lesssim 0.079$ the
infection free equilibrium loses it's stability and an endemic equilibrium
is born; for $p\gtrsim 0.079$ the endemic equilibrium loses it's stability
through a Hopf bifurcation and a stable limit cycle is born. See the upper
dashed Hopf bifurcation curve in Fig.~\ref{f:0st_Hopf}(b). In the case of $%
K=12 $, the infection free equilibrium is unstable for all values of $p$.
However, the endemic equilibrium is stable for $0\leq p\lesssim 0.011$. When 
$p\gtrsim 0.011$, both equilibria are unstable and instead a stable limit
cycle is observed. See the lower solid Hopf bifurcation curve in Fig.~\ref%
{f:0st_Hopf}(b).

As in the network model, the disease goes completely extinct for relatively
large $p$ values (for $K=12$, $p_{c}=0.43$ and for $K=8$, $p_{c}=0.87$),
despite the fact that the infection free equilibrium is unstable (see Fig.~%
\ref{f:de}(c)). The extinction should be attributed to the synchronization
effect that takes place for these high $p$ values. That is, the dynamics are
such that a large proportion of the population become infected together and
proceed on to move to the recovered class together. The synchronization
requires the initiation of a strong epidemic, implying that $R_{0}$ must be
greater than unity, which explains why the infection free equilibrium is
unstable in this regime. Nevertheless the disease becomes extinct due to the
synchronization effect.

\subsection{Vaccination}

A \emph{random vaccination} scheme may be incorporated into the difference
equation model by replacing $S_{t}$ in Eqn.~(\ref{e:Itp1}) with $(1-v)S_{t}$%
. The parameter $v$ is the proportion of susceptibles vaccinated per time
unit. Denote by $v_{e}$ the threshold proportion of vaccinated needed for
disease extinction. A simple algebraic expression can be obtained for the
extinction threshold by linearizing Eqn.~(\ref{e:Itp1})  about the infection
free equilibrium. Then, by using equation (\ref{e:bifcrv0}), disease
extinction is reached if: 
\begin{equation}
v\geq v_{e}=1-\frac{1}{R_{0}} .  \label{e:vacp2}
\end{equation}
For the parameter values of Figs.~\ref{f:de} and \ref{f:0st_Hopf}, the
vaccination threshold is $v_{e}=0.5297$. Numerical simulations corroborate
the existence of this threshold. 

As the difference equation does not give any information regarding spatial
patterns, it is impossible to apply the spatially oriented targeted
vaccination scheme described above. However, targeted vaccination schemes
may nevertheless be explored by differentiating between vaccinating nearest
neighbors and short-cuts. This can be achieved by replacing the term $I_{t}$
in equations (\ref{e:St}) and (\ref{e:Itp1}) with the term $(1-v_{K})I_{t}$
for nearest neighbors and $(1-v_{p})I_{t}$ for short-cuts. The results
reveal that the extinction threshold is very sensitive to and lowered
dramatically by the term $v_{p}$ for vaccinating short-cuts, while the term
for vaccinating nearest neighbors, $v_{K}$, has little influence. Although
according to the model the infection free equilibrium is never stable for
the syphilis disease in a small world type society, it appears that targeted
vaccination effectively reduces the number of the actual contacts of the key
individuals, thereby reducing $R_{0}<1$ in the vaccinated population. A
detailed study of the targeted vaccination is left for future work.

\section{Conclusion and discussion}

Two models for studying the dynamics of diseases with temporary immunity in
complex population networks have been proposed -- a lattice model, which
incorporates spatial information, and a difference equation model, which
allows an analytic approach. The study focuses on the example of syphilis
epidemics, which is a representative STD targeted to be eliminated in the US
with little success so far. The network model reveals that diseases with
temporary immunity on a small world contact network exhibit periodicity and
waves of epidemics, some of which become pacemaker centers. It is shown that
by eliminating pacemakers through vaccination, the disease goes to
extinction within $1-2$ periods, where only about $20\%$ of the population
requires vaccination. This is in contrast to standard vaccination programs
that set out to achieve herd immunity by vaccinating over $80\%$ of the
population. Moreover, by treating only those individuals in high risk
pacemaker areas, it minimizes the application of the vaccination with its
possible risks to the larger population. The difference equation model
allows further investigation of the Hopf bifurcation lying at the heart of
the pacemaker phenomena. The models presented here were constructed in
accordance to the US syphilis datasets (see \cite{CDC, CDC2005} and \cite%
{GFG2005}). The two models complement each other, allowing a more profound
view of the dynamics.

This work addresses the controversy as to whether syphilis epidemics recur
approximately every ten years due to the temporary immunity it endows to
infected individuals (\cite{GFG2005}) or due to changing patterns in social
behavior (\cite{CDC, CDC2005}). As shown here, both factors are crucial for
recurrent syphilis epidemics. Thus, for example, oscillations cannot occur
outside the small world regime even in the presence of strong temporary
immunity. For zero or very small $p$ values (corresponding to none, or a
very few, short-cut links), an epidemics cannot develop. Moreover, the
analysis performed on the difference equation model reveals that if \emph{all%
} individuals in a small world type population network have only a small
number of contacts, the infection free equilibrium is stable. In addition,
on the other side of the scale, it is pointed out that for large enough $p$
outside the small world region (corresponding to many short-cut links) a
synchronization effect takes place, eradicating the epidemics due to
exhaustion of susceptibles. In contrast, \emph{a society whose social
behavior approximates a small world network with moderate heterogeneous
levels of promiscuity would sustain the periodic recurrences of the syphilis
epidemics approximately every ten years}.

Finally, we conjecture that complete disease extinction is nevertheless
achievable by a targeted vaccination scheme similar to that presented here.
The targeting and vaccination of key individuals, effectively reduces $R_{0}$
to less than unity in the vaccinated population, thereby leading to disease
extinction.

\appendix

\section{Appendix1: Technical details for the difference equation model}

Here we present the technical details of the stability analysis performed
for the difference equation model (\ref{e:Itp1}). Consider the \textit{S}%
usceptible - \textit{I}nfectious - \textit{R}ecovered - \textit{S}usceptible
(SIRS) dynamics. Assume $\tau _{I}$ is the amount of time units an
individual spends in the Infectious class and that $\tau _{R}$ represents
the time units an individual later spends in the Recovered class. As for
syphilis $\tau _{I}=1$ (where a time unit represents $6$ months), the
proportion of recovered individuals can be described by the sum: $\Sigma
_{i=1}^{\tau _{0}-1}I_{t-i}$. Thus, denoting by $S_{t}$ the proportion of
susceptibles in the population at time $t$, we obtain:

\begin{equation}
S_{t}=1-I_{t}-\Sigma _{i=1}^{\tau _{0}-1}I_{t-i},  \label{ea:St}
\end{equation}%
where $\tau _{0}=\tau _{I}+\tau _{R}$. Set $I_{t}$ to be the proportion of
infectives at time $t$, denote by $K$ the number of connections an
individual has on average and let $p$ be the proportion of short-cut links
an individual has among it's $K$ connections. Then, denote by $K_{t}^{\inf K}
$ the number of infectious \emph{nearest neighbors }an individual node has
at time $t$ and by $K_{t}^{\inf p}$ the number of infectious contacts via 
\emph{short-cut links} an individual node has among its $K$ connections at
time $t$. Then:

\begin{eqnarray}
K_{t}^{\inf K} &=&I_{t}(1-p)K,  \label{ea:KinfK} \\
K_{t}^{\inf p} &=&I_{t}pK.  \label{ea:Kinfp}
\end{eqnarray}%
Set $q_{p}$ as the probability of being infected via a short-cut link and
set $q_{K}$ as the probability of being infected by a nearest neighbor,
where $q_{K}<q_{p}<1$. Hence, the $SIRS$ dynamics on a network with nearest
neighbors and short-cut connections can be described by: 
\begin{equation}
I_{t+1}=(1-(1-q_{K})^{K_{t}^{\inf K}}(1-q_{p})^{K_{t}^{\inf p}})S_{t},
\label{ea:Itp1}
\end{equation}%
The dynamics of (\ref{ea:Itp1}) are at equilibrium for the solutions, $%
I^{\ast }$, of the equation (\ref{ea:defp}):

\begin{equation}
I^{\ast }=(1-(1-q_{K})^{I^{\ast }(1-p)K}(1-q_{p})^{I^{\ast }pK})(1-\tau
_{0}I^{\ast }).  \label{ea:defp}
\end{equation}%
It is easily seen that the infection free equilibrium $I^{\ast }=0$ is
always a solution and that an endemic equilibrium $I^{\ast }>0$ is a
solution for most parameters relevant for syphilis.

Substituting $J_{t}=I_{t}-I^{\ast }$ and linearizing Eqn.(\ref{ea:Itp1})
about $I^{\ast }$, we obtain: 
\begin{equation*}
J_{t+1}=(1-q_{K})^{(1-p)KI^{\ast }}(1-q_{p})^{pKI^{\ast }}\left(
R_{0}(1-\tau _{0})J_{t}+\Sigma _{i=0}^{\tau _{0}-1}I_{t-i}\right) -\Sigma
_{i=0}^{\tau _{0}-1}I_{t-i}+h.o.t.,
\end{equation*}%
where, 
\begin{equation*}
R_{0}=-K((1-p)\ln (1-q_{K})+p\ln (1-q_{p})).
\end{equation*}

Now substitute $J_{t}=J_{0}\lambda ^{t}$ to obtain the characteristic
polynomial, 
\begin{equation}
\lambda ^{\tau _{0}}+\alpha \lambda ^{\tau _{0}-1}+\beta \lambda ^{\tau
_{0}-2}+\cdot \cdot \cdot +\beta \lambda +\beta =0,  \label{ea:char_poly}
\end{equation}%
where: 
\begin{eqnarray}
\alpha &=&(1-q_{K})^{(1-p)KI^{\ast }}(1-q_{p})^{pKI^{\ast }}\left(
R_{0}(\tau _{0}I^{\ast }-1)-1\right) +1,  \label{ea:ab} \\
\beta &=&1-(1-q_{K})^{(1-p)KI^{\ast }}(1-q_{p})^{pKI^{\ast }}.  \notag
\end{eqnarray}%
The stability of the infection free equilibrium is derived from substituting 
$I^{\ast }=0$ and requiring that $\left\vert \lambda \right\vert =R_{0}<1$.
The stability analysis of the infection free equilibrium is presented in the
main text (and Fig.~\ref{f:0st_Hopf}(a)). Here we provide details regarding the endemic
equilibrium stability and the calculation of the Hopf bifurcation curve(s),
presented in Fig.~\ref{f:0st_Hopf} in the main text. This calculation is inspired by a
calculation of the Hopf bifurcation of a mean field difference equation
model in \cite{difeq}.

Assume such a bifurcation exists and substitute $\lambda =e^{i\phi }$ (as
stability changes at $\left\vert \lambda \right\vert =1$) into Eqn.(\ref%
{ea:char_poly}). Separating real and imaginary parts we obtain two
equations: 
\begin{eqnarray}
\alpha  &=&-\cos (\phi )-\cot \left( \frac{\phi \tau _{0}}{2}\right) \sin
(\phi ),  \label{ea:abf} \\
\beta  &=&\frac{2\csc \left( \frac{\phi \tau _{0}}{2}\right) \sin \left( 
\frac{\phi }{2}\right) \sin (\phi )}{\cos \left( \frac{\phi (\tau _{0}-2)}{2}%
\right) -\cos \left( \frac{\phi \tau _{0}}{2}\right) }.  \notag
\end{eqnarray}%
Substituting Eqns.(\ref{ea:ab}) into Eqns.(\ref{ea:abf}) and adding Eqn.(\ref%
{ea:defp}) results in a system of three equations in the seven variables: $%
q_{K},q_{p},p,K,I^{\ast },\tau _{0},\phi $. Some of these parameters can be
fixed to values relevant for syphilis: $\tau _{0}=11$, $K=8$ or $12$ and $%
q_{K}=0.11$. $\phi $ can be viewed as the frequency of the periodic solution
emerging at the bifurcation, where the period of the limit cycle (when
exists) is $T\thickapprox \frac{2\pi }{\phi }$. As for the syphilis
parameter values the period of oscillation is $T\thickapprox 20$ time units,
where $1$ time unit $=0.5y=6$ months, we fix $\phi =0.3$. Now, with these
fixed parameter values, we use the Newton method to solve Eqns.(\ref{ea:abf}%
) and (\ref{ea:defp}), using the syphilis parameter values as the initial
guess, once for $K=8$ (dashed curve in Fig.~\ref{f:0st_Hopf}(b)) and once for $K=12$ (solid
curve in Fig.~\ref{f:0st_Hopf}(b)). The results are plotted in the $(p,q_{p})$ plane (see
Fig.~\ref{f:0st_Hopf}(b) \ in the main text). In Fig.~\ref{f:0st_Hopf}(b), the two bifurcation curves (one
for $K=8$ \ - dashed and one for $K=12$ - solid) are presented, where below
each curve an endemic equilibrium $I^{\ast }>0$ is stable. At the
bifurcation curve the equilibrium loses its stability and a stable limit
cycle is born so that above the curve a stable limit cycle coexists with two
repelling (unstable) equilibria - the infection free and an endemic.

\end{document}